\documentclass[a4paper,10pt]{article}
\usepackage{enumerate}
\usepackage{color}
\usepackage[utf8]{inputenc} 
\usepackage[english]{babel}
\usepackage[T1]{fontenc}
\usepackage{graphicx}
\usepackage{amsfonts,amssymb,amsmath,latexsym,amsthm}
\usepackage{textcomp}
\usepackage[pdftex]{hyperref}
\usepackage{geometry}
\DeclareMathOperator{\sech}{sech}

\title{Soliton interactions with an external forcing: the modified Korteweg-de Vries framework}
\author{Marcelo V. Flamarion$^{1}$ and Efim Pelinovsky$^{2,3}$}
\date{}

\begin{document}
\maketitle
\begin{center}
{\footnotesize $^1$Unidade Acad{\^ e}mica do Cabo de Santo Agostinho, \\
UFRPE/Rural Federal University of Pernambuco, BR 101 Sul, Cabo de Santo Agostinho-PE, Brazil,  54503-900 \\
marcelo.flamarion@ufrpe.br }

\vspace{0.3cm}
{\footnotesize $^{2}$Institute of Applied Physics, 46 Uljanov Str., Nizhny Novgorod 603155, Russia. \\
 $^{3}$National Research University--Higher School of Economics, Moscow, Russia. }

%{\footnotesize ORCID Number: 0000-0002-0877-4831}

\end{center}

%\maketitle

\begin{abstract} 
The aim of this work is to study asymptotically and numerically the interaction of solitons with an external forcing with variable speed using the forced modified Korteweg-de Vries equation (mKdV). We show that the asymptotic predictions agree well with numerical solutions for forcing with constant speed and linear variable speed. Regarding forcing with  linear variable speed, we find regimes in which the solitons are trapped at the external forcing and its amplitude increases or decreases in time depending on whether the forcing accelerates or decelerates. 

	\end{abstract}

\section{Introduction}
The forced Korteweg-de Vries equation (fKdV) is broadly used to describe the propagation of waves of small amplitudes produced by moving sources with constant speed and small amplitudes  \cite{Akylas:1984, Wu:1982, Wu:1987, Wu1, Milewski:2004, Grimshaw:1986, Grimshaw:2016, Grimshaw:2019, Choi:2008, Ermakov:2019}. For instance this equation has been used to investigate flows past obstacles in hydrodynamics, ship wakes, trapped waves \cite{Malomed:1993, Grimshaw:1993, Grimshaw:1994, Lee:2018, Kim:2018, LeeWhang:2018, Flamarion-RibeiroJr:2021},  internal waves in stratified fluid flows and so on \cite{Baines:1995}. In many practical applications, for example, ship wakes or atmospheric flows,  the speed of the external force may vary along the time or depend on the medium. Consequently, it can produce surface or internal solitary waves whose speed depend on the medium in which these waves propagate. For an external force with a time-dependent speed,  a complete asymptotic and numerical study of the fKdV equation was first given by Grimshaw et al. \cite{Grimshaw:1996}. The authors showed a good agreement of results predicted by both theories and conditions for trapping were determined within the asymptotic framework.  Similar studies in the in the context of gravity-capillary waves were done later by Flamarion \cite{Flamarion:2022a}  and in the non-integrable Whitham equation in  \cite{Flamarion:2022b}.

When waves have larger amplitude or the nonlinear coefficient is small, nonlinearity is dominant  in the dynamic and the fKdV equation fails to predict many phenomena. In that case, nonlinear terms of higher-order have to be taken into account and  the forced modified KdV equation (mKdV), which incorporates a cubic nonlinearity, arises. Among the problems that can be investigated in this framework, we mention waves on the surface of conductive fluid in an electric field \cite{Perelman:1974}, wave in quantum-dimensional films, elastic waves in solids \cite{Pavlov:1998}.  Although the mKdV equation is integrable, the nonlinear dynamics is more complicated than the fKdV equation and the sign of the cubic nonlinearity plays a fundamental role on the qualitative behaviour of the solutions. For instance, when cubic nonlinearity is positive, the mKdV equation admits breathers (traveling oscillating  moving wave packets) and solitons of both polarities as solutions. This equation has been widely used to study breathers and solitons with different polarities \cite{Pelinovsky:2002, Pelinovsky:2020}. In particular, an asymptotic study on trapped waves in this framework was done by Pelinovsky \cite{Pelinovsky:2002} and the results were compared qualitatively with the fKdV equation. Asymptotic results predict that trapped waves can occur only when the forcing and soliton have the same polarity, however fully numerical results were not reported.

The goal of this article is to investigate numerically and asymptotically the interaction of solitons with an external forcing with variable speed using the forced modified Korteweg-de Vries equation with positive cubic nonlinearity. Asymptotic and numerical results are compared qualitatively and quantitatively and conditions of trapping are discussed. In particular, we show that for forcing with linear variable speed, it is possible to trap  a soliton at the external forcing and that  its amplitude increases or decreases in time depending on whether the external force accelerates or decelerates. 

This article is organized as follows. In section 2 we present the mathematical formulation of the problem. Asymptotic results  are presented in section 3 and  numerical results in section 4. The conclusion is presented in section 5.

\section{The forced modified Korteweg-de Vries equation}
We consider the modified Korteweg-de Vries equation with a forcing term of variable speed ($v(t)$) in canonical form
\begin{equation}\label{mfKdV1}
u_{t} +cu_x+6u^{2} u_x+u_{xxx}=\epsilon f_{x}\Big(x-\int v(t) dt\Big).
\end{equation}
to investigate the interaction of solitons with a external forcing field. Here, we denote by $u(x,t)$  the wave field, $f(x)$  the external forcing that travels with variable speed $v(t)$, $c$ the long-wave propagation speed and $\epsilon$  a small positive parameter.  It is convenient to rewrite equation (\ref{mfKdV1}) in the forcing moving frame.  To this end, we write the traveling variables $$x'=x-\int v(t) dt, \;\ t'=t.$$
In the new coordinate system, dropping the primes equation (\ref{mfKdV1}) reads
\begin{equation}\label{mfKdV}
u_{t} +\Delta(t)u_x+ 6u^{2} u_x+u_{xxx}=\epsilon f_{x}(x),
\end{equation}
where 
\begin{equation} \label{deviation}
\Delta(t)=c-v(t)
\end{equation}
is the deviation speed. This important parameter measures the difference between the linear long-wave speed and the speed of the external forcing \cite{Grimshaw:1996}.
Assuming that the forcing vanishes at infinity, equation (\ref{mfKdV}) conserves the total mass $(M(t))$, with
\begin{equation}\label{mass}
\frac{dM}{dt}=0, \mbox {where } M(t)=\int_{-\infty}^{\infty}u(x,t)dx.
\end{equation} 
Additionally, the rate of change of momentum $(P(t))$ is balanced by the external forcing through the equation
\begin{equation}\label{momentum}
\frac{dP}{dt}=\int_{-\infty}^{\infty}u(x,t)\frac{df(x)}{dx}dx, \mbox {where } P(t)=\frac{1}{2}\int_{-\infty}^{\infty}u^{2}(x,t)dx.
\end{equation} 
In the absence of an external forcing, the mKdV admits  solitons as solutions  \cite{Pelinovsky:2002}, which are given by the expressions
\begin{equation}\label{solitary}
u(x,t)=a\sech(a\mathbf{\Phi}), \mbox{  where }  \;\ \mathbf{\Phi}=x-qt-x_0, \;\ q = c-v+a^2,
\end{equation}
where $a$ is amplitude of the soliton,  $x_0$ is its the initial position of the crest. Here, $a$ is allowed to be negative, which represents depression solitary waves.

\section{Asymptotic results}
Asymptotic results on the interaction of a solitons with an external forcing were first  reported  Pelinovsky \cite{Pelinovsky:2002} for elevation waves. Here, we present his main results and additionally we expand formally his analysis to depression solitons and solutions of dynamical systems are obtained numerically. With this in mind, we assume that the wave field is close to the soliton, but the parameters  vary slowly in time. The soliton is described by the formulas
\begin{equation}\label{solitary}
u(\mathbf{\Phi},T)=a(T)\sech(a(T)\mathbf{\Phi}), \mbox{  where }  \;\ \mathbf{\Phi}=x-\mathbf{\Psi}(T), \mbox{ and } \mathbf{\Psi}(T) =x_0+\frac{1}{\epsilon}\int q(T)dT,
\end{equation}
where functions $a$ and $q$ are determined from the interaction between the wave field and the external field. We  formally introduce the ``slow time" by considering the new variable $T=\epsilon t$. Assuming a weak force ($\epsilon\ll 1$), we seek for a solution in the form of the asymptotic expansion 
\begin{align} \label{Asymptotic}
\begin{split}
& u(\mathbf{\Phi},T)=u_{0}+\epsilon u_1+\epsilon^{2} u_2+\cdots , \\
& q(t) = q_0 + \epsilon q_1 + \epsilon^{2} q_2+\cdots. \\
\end{split}
\end{align}
At the lowest order of the perturbation theory, the solutions $u_0$ and $q_0$  are defined as in equation (\ref{solitary}) with the position of the initial crest given by $x_0=c-v-a^2$.

The momentum balance equation (\ref{momentum}) and kinematic condition yields the dynamical system
\begin{align} \label{DS0}
\begin{split}
& \frac{d\mathbf{\Psi}}{dt} = a\int_{-\infty}^{\infty}\sech(a\mathbf{\Phi})\frac{df}{d\mathbf{\Phi}}(\mathbf{\Phi}+\mathbf{\Psi})d\mathbf{\Phi}, \\
& \frac{da}{dt}=\Delta(T)+a^2, 
\end{split}
\end{align}
If the forcing is broad in comparison with the soliton length, the soliton can be considered as a delta-function and
consequently the dynamical system for the amplitude and position of the crest becomes
\begin{align} \label{DS}
\begin{split}
& \frac{dx}{dt} = \Delta(t)+a^2, \\
& \frac{da}{dt}=\pi\frac{df(x)}{dx}.
\end{split}
\end{align}
The external forcing profile is chosen as
\begin{equation}\label{impulse}
f(x)=b\exp{(-x^2/L^2)},
\end{equation}
where $L$ and $b$ are constants. Notice that if $u(x,t)$ is a solution of equation (\ref{mfKdV}) for a forcing term $f(x)$, then $-u(x,t)$ is a solution for the same problem with forcing term $-f(x)$. Therefore,  we restrict ourselves to consider only the case $b>0$ and similar results are valid for $b<0$.

\subsection{Soliton interactions with an external forcing with constant speed}
\begin{figure}[h!]
	\centering	
	\includegraphics[scale =1.1]{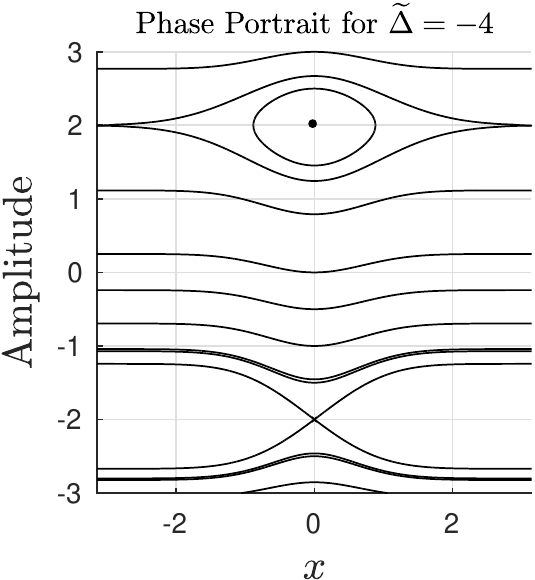}
 	\includegraphics[scale =1.1]{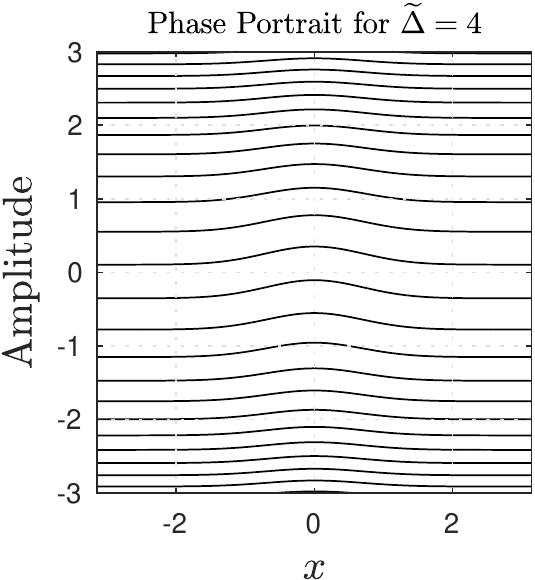}
	\includegraphics[scale =1.1]{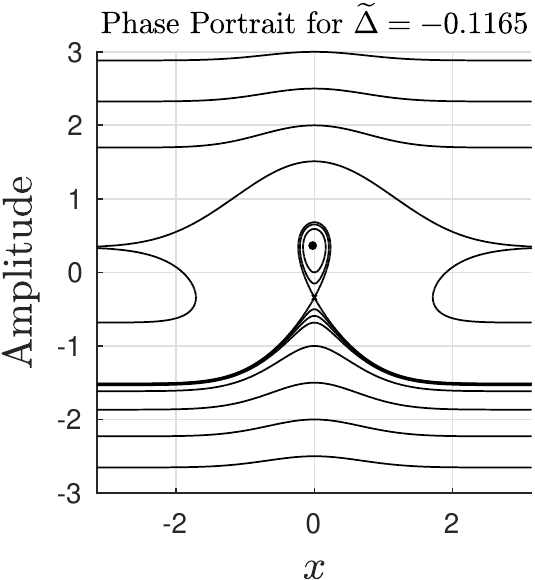}
	\includegraphics[scale =1.1]{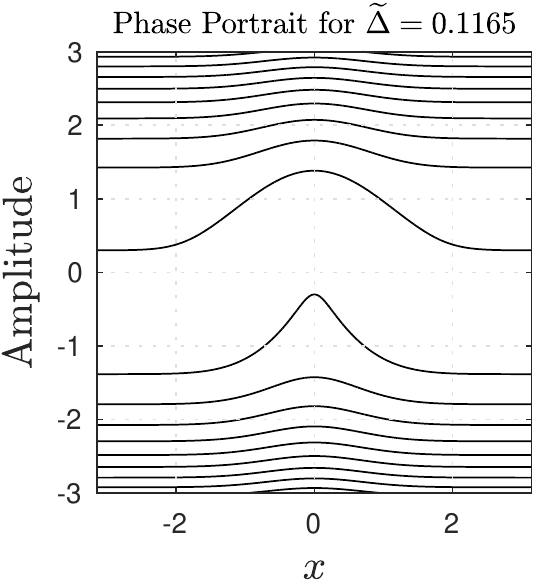}
	\caption{Phase portraits for the dynamical system (\ref{DS}). Circles correspond to centres and  crosses to saddles.}
	\label{Fig1}
\end{figure}
When the forcing moves at constant speed $v$, solutions of the dynamical system (\ref{DS}) has two equilibrium points located at the maximum of the forcing and the amplitude of the resonance soliton can be obtained explicitly as 
\begin{equation} \label{equilibrium}
a_{0}^{\pm}=\pm\sqrt{-\Delta}.
\end{equation}
Such a soliton exists only if the external disturbance travels faster than a long-wave  $(v>c)$. The equilibrium position is a centre if the disturbance and the soliton have the same polarity and a saddle otherwise. In the later, a soliton moves away from the external field without reversing its direction and its interaction ceases once the waves passes over the external forcing. Solutions of system (\ref{DS}) are represented by streamlines i.e., solutions are the level curves of the stream function $H(x,a)$, which is given by
\begin{equation}\label{streamfunction}
H(x,a) = -\pi f(x)+\Delta a+\frac{a^3}{3}.
\end{equation}

In order to analyse the phase portrait of system (\ref{DS})  for the external forcing (\ref{impulse}) we rescale the variables as follows: the coordinate $x$ is rescaled with respect to $L$, and the amplitude $a$ with respect to $(b\pi)^{1/3}$, where $b>0$ is the amplitude of the external forcing. This yields the parameter 
\begin{equation}
\widetilde{\Delta}= \frac{\Delta}{(\pi b)^{2/3}}.
\end{equation}
In this new coordinates system, the  streamfunction reads
\begin{equation}\label{streamfunction}
H(x,a) = -e^{-x^2}+\widetilde{\Delta} a+\frac{a^3}{3}.
\end{equation}

Figure \ref{Fig1} displays typical phase portraits of system (\ref{DS}) for different values of the parameter $\widetilde{\Delta}$. Notice that trapped solitons do not occur for positive values of $\widetilde{\Delta}$ (see Figure \ref{Fig1}  right). On the other hand,  for large negative values of $\widetilde{\Delta}$ in we have a  trapped elevation solitons close to the centre. As  $\widetilde{\Delta}$ remains negative and approaches zero, both equilibrium points approach the axis $a=0$. Consequently, the dynamical system predicts the existence of trapped solitons whose polarity change over time.  It is worth to mention that the asymptotic theory fails for solitons of small amplitudes. It occurs because the soliton width unrestrictedly increases. Additionally, a soliton cannot  change its polarity in the asymptotic theory. When the soliton amplitude tends to zero, all terms in the forced mKdV equation have the same order consequently perturbation theory breaks. It is also follows from equation (\ref{momentum}) that  the polarity of the soliton is conserved.

Although we do not show here,  a bifurcation takes place when the centre and saddle coalesce into a single point for a small value of $\widetilde{\Delta}$. We do not present corresponding figure because as we mentioned above the asymptotic theory fails for solitons of small amplitudes.

\subsection{Soliton interactions with an external forcing with linear variable speed}
\begin{figure}[h!]
	\centering	
	\includegraphics[scale =1.1]{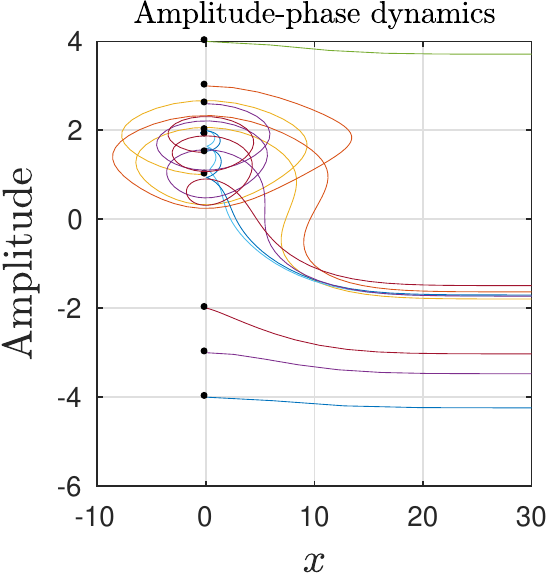}
 	\includegraphics[scale =1.1]{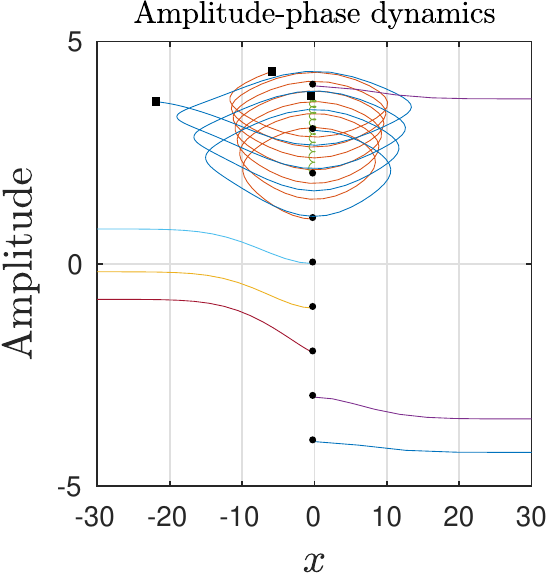}
	\caption{Amplitude-phase dynamics for different initial values of (\ref{DS}) and deviation speed (\ref{lineardeviation}). Parameters: $\Delta_0=-4$,   $g=-0.1$ (left) and $g=0.1$ (right). Circles correspond to the starting points and squares to the ending points.}
	\label{Fig2}
\end{figure}
\begin{figure}[h!]
	\centering	
	\includegraphics[scale =1.1]{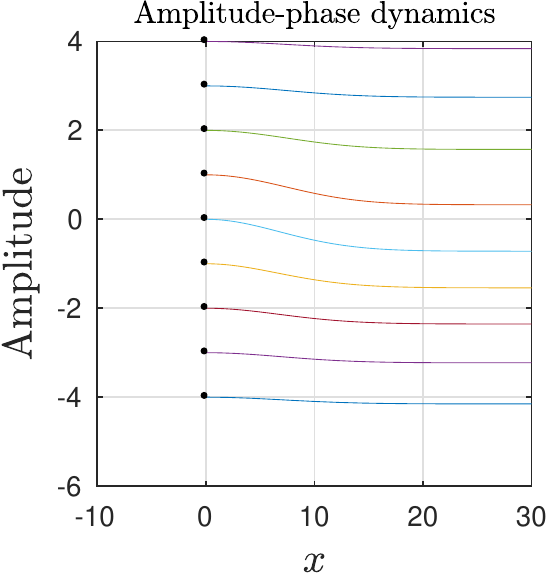}
 	\includegraphics[scale =1.1]{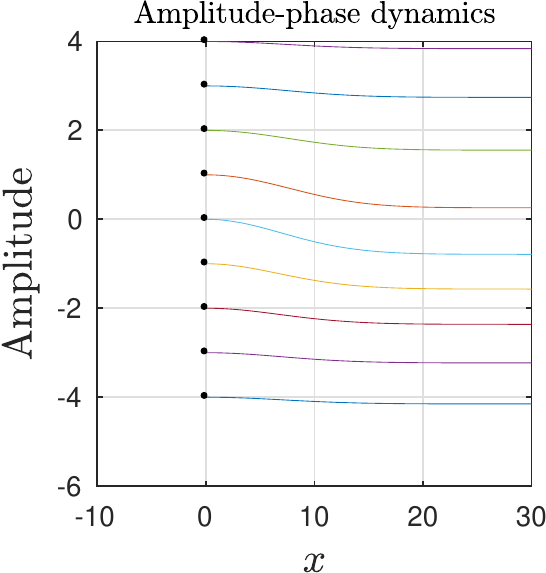}
	\caption{Amplitude-phase dynamics for different initial values of (\ref{DS}) and deviation speed (\ref{lineardeviation}). Parameters: $\Delta_0=4$,   $g=-0.1$ (left) and $g=0.1$ (right). Circles correspond to the starting points.}
	\label{Fig2b}
\end{figure}
In this section we consider the external forcing with variable speed. More precisely, we focus on  forcings with constant acceleration. In this case,  the variable speed is defined as 
\begin{equation}
v(t) = v_{0} + gt,
\end{equation}
where $v_0$ is the initial speed of the forcing and $g$ its acceleration. The speed deviation is defined as
\begin{equation}\label{lineardeviation}
\Delta(t) = c-v(t)= \Delta_{0} - gt.
\end{equation}
Although we consider a forcing with linear variable speed, solutions of (\ref{DS}) cannot be obtained in closed form, thus we solve the dynamical system (\ref{DS}) numerically. We focus on the case that the accelaration ($g$) is small.  The idea behind choosing $g$ small lies on the fact that  the dynamical system (\ref{DS}) is "nearly" autonomous. This way we can consider $g$ as a small perturbation of the autonomous system (\ref{DS})  and investigate the behaviour in the neighbourhood of the equilibrium points.   We recall that no crossing through the axis $a=0$ is allowed in the asymptotic theory. However, we can still extract qualitative results from the solutions of system (\ref{DS}) and later compare them with the fully numerical simulations.

In order to give a good description of the amplitude-phase dynamics for the non autonomous system, we compute solutions numerically for different initial data. In particular, we consider initial data with amplitudes close to the equilibrium points (\ref{equilibrium}) of the autonomous dynamical system ($g=0$). The amplitude-phase dynamics  is displayed in great details in Figure \ref{Fig2}. The circles represent the initial position of the soliton and its initial amplitude.  Notice that close to the centre of the autonomous dynamical system,  a soliton bounces back and forth at the external forcing while its amplitude decays with time as the forcing decelerates ($g<0$)  and increases with time as the forcing  accelerates ($g>0$). As the initial data is taken far from the centre of the autonomous system the soliton moves pass the external force and travels with constant speed. Close to the saddle of the autonomous system, a soliton moves away from the external field without reversing its direction. On the other hand, when the deviation speed is positive solitons are no longer trapped. All solitons move past the external field without reversing direction. This can be seen in the amplitude-phase dynamics depicted in Figure \ref{Fig2b}.

\section{Numerical results}
Solutions of equation (\ref{mfKdV}) are computed numerically through the standard pseudospectral  method \cite{Trefethen:2000}. The computational domain $[-L,L]$ is discretised with $N$ points  uniformly spaced.  Spatial derivatives are computed spectrally and to prevent spatial effects of the periodicity  the computational domain is taken  sufficiently large. The time evolution is calculated through the Runge-Kutta fourth-order method with time step $\Delta t$. Typical computations are performed using $N=2^{12}$ Fourier modes with $L=512$ and $\Delta t=10^{-3}$. A study of the resolution of a similar numerical method for the forced KdV equation can be seen in \cite{Marcelo-Paul-Andre}. Additionally, in the absence of the external forcing, we verified that the numerical method conserves mass and  momentum. For solitons of form (\ref{solitary}) with amplitude in the range of the interval $[-0.8,0.8]$, we have that the relative error is at least order
$$\frac{\displaystyle\max_{0\le t\le 10^{4}}|M(t)-M(0)|}{|M(0)|}=\mathcal{O}(10^{-16}).$$
Besides, we also verify that 
	$$\frac{\displaystyle\max_{0\le t\le  10^{4}}|P(t)-P(0)|}{|P(0)|}=\mathcal{O}(10^{-12}).$$

We start our discussion comparing the results predicted by the asymptotic theory with the fully numerical computations for external forcings with constant speed. More precisely, we investigate if the centre points of the dynamical system (\ref{DS}) define trapped waves for equation (\ref{mfKdV}). Since there is a long list of parameters to be considered in the study of the interaction between a soliton and an external forcing, we fix a few parameters, namely, $\epsilon=0.01$, $L=10$. %Although we do not have a phase portrait for equation (\ref{mfKdV}), solutions in the space amplitude vs. crest position can still be plotted. This is interesting because allow us to verify whether the predictions of the asymptotic is valid or not. 

\subsection{Soliton interactions with an external forcing with constant speed}

\begin{figure}[h!]
	\centering	
	\includegraphics[scale =1]{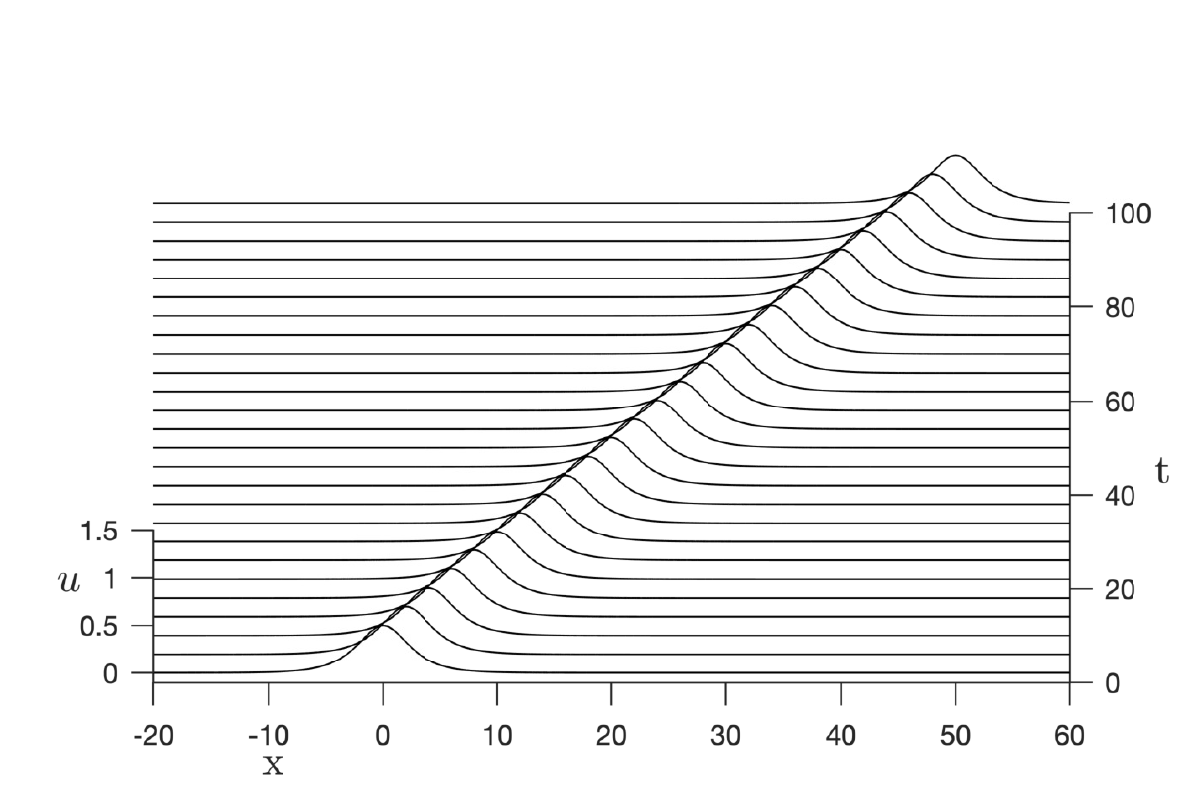}
 	\includegraphics[scale =1]{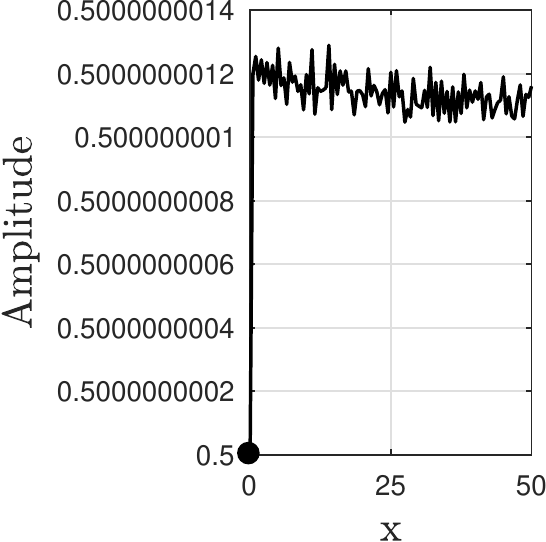}
         \includegraphics[scale =1]{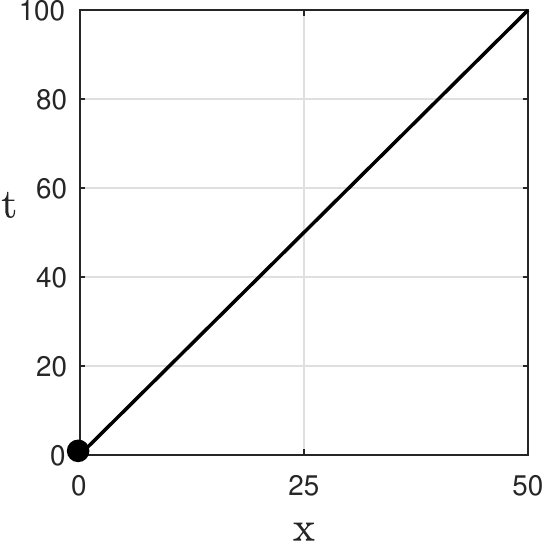}
	\caption{Top (left): nontrapped soliton with the same polarity of the forcing. Bottom (left): the space amplitude vs. crest position. Bottom (right): the crest position along the  time. Parameters: $a=0.5$, $\Delta = a^2$ and $b=1$.}
	\label{Fig6}
\end{figure}

The asymptotic theory predicts non-trapping solitons for $\Delta>0$. In order to verify these predictions for the forced mKdV equation, we run a large number of simulations and indeed all solitons move past the external force without changing directions. The amplitude of the soliton varies slightly as long as there is an interaction with the external forcing. Once the interaction ceases, the soliton travels with constant amplitude and speed (see Figure  \ref{Fig6} (bottom)).  This typical behaviour is illustrated in Figure \ref{Fig6}. The soliton is affected by the presence of the external force, but at large times it travels with constant speed preserving its shape.

\begin{figure}[h!]
	\centering	
	\includegraphics[scale =1]{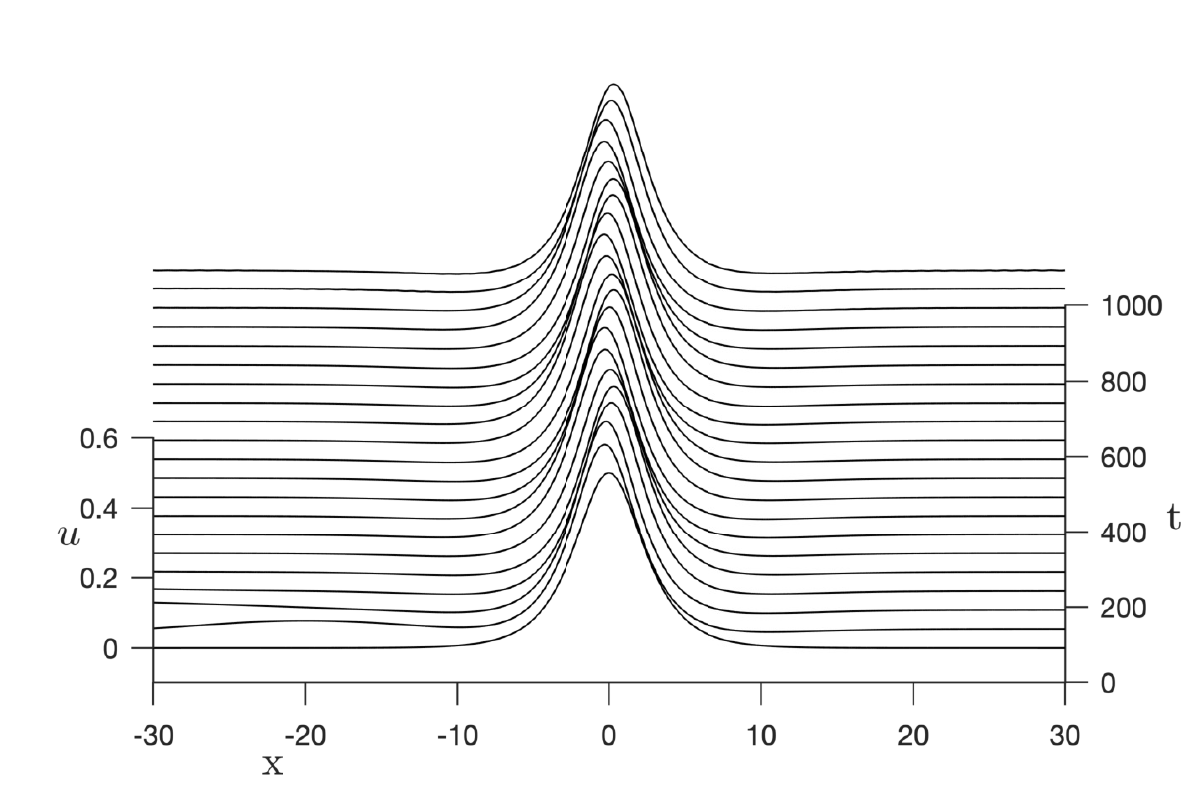}
 	\includegraphics[scale =1]{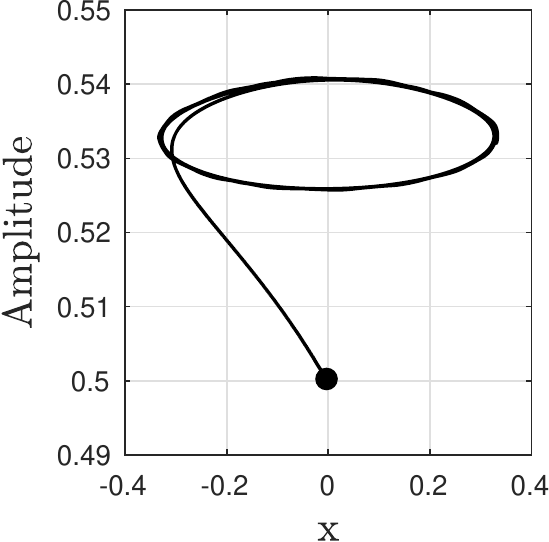}
         \includegraphics[scale =1]{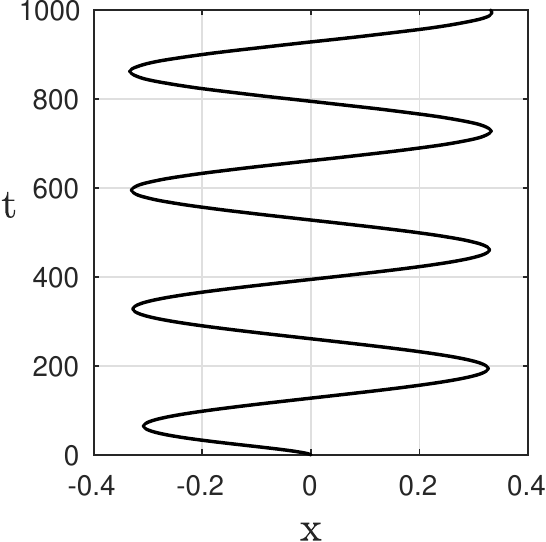}
	\caption{Top (left): trapped soliton with the same polarity of the forcing. Bottom (left): the space amplitude vs. crest position. Bottom (right): the crest position along the  time. Parameters: $a=0.5$, $\Delta = -0.1165$ and $b=1$.}
	\label{Fig3}
\end{figure}

The more interesting case occurs when $\Delta<0$ because solitons can be trapped at the external field. The asymptotic theory predictions agree well with the numerical solutions qualitatively and quantitatively. Notice that the asymptotic solution is truncated at order $\mathcal{O}(\epsilon)$, thus this is the error expected when comparing the numerical solutions with the asymptotic ones. Figure \ref{Fig3} displays a typical solution of equation (\ref{mfKdV}) for an initial soliton with the amplitude and crest position chosen as a centre of the dynamical system (\ref{DS}). Notice that the crests remain confined in a small region for large values of time, i.e., in space amplitude vs. crest position trajectories are almost closed. Additionally, in the space  position of crest vs. time,  the soliton behaviour resembles  an harmonic oscillator. Although we do not show here, similar results also occur for different choices of the initial amplitude and crest-position of the soliton close to the centre point predicted by the asymptotic theory. 

\subsection{Soliton interactions with an external forcing with linear variable speed}

This section is devoted to the study of the interaction between solitons and an external force with linear variable speed. This case is of relevance because it is the first-order approximation of a smooth variable speed. 

\begin{figure}[h!]
	\centering	
	\includegraphics[scale =1]{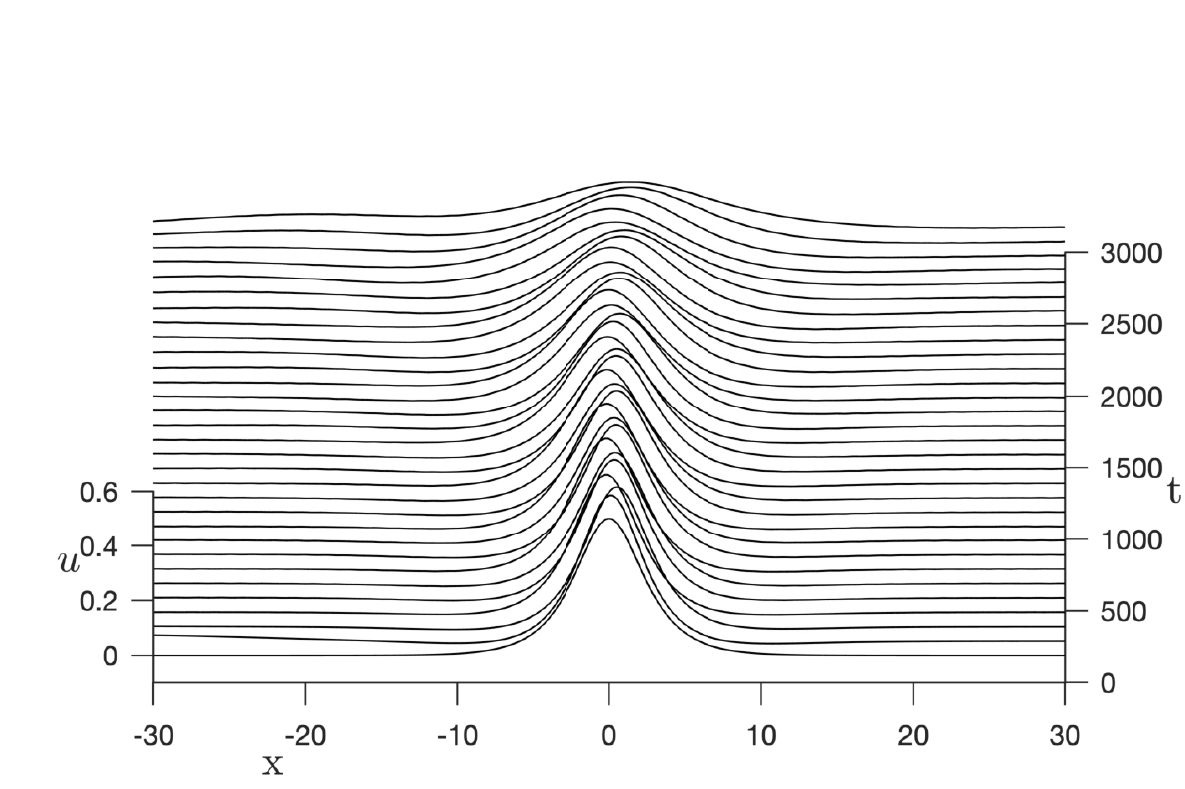}
 	\includegraphics[scale =1]{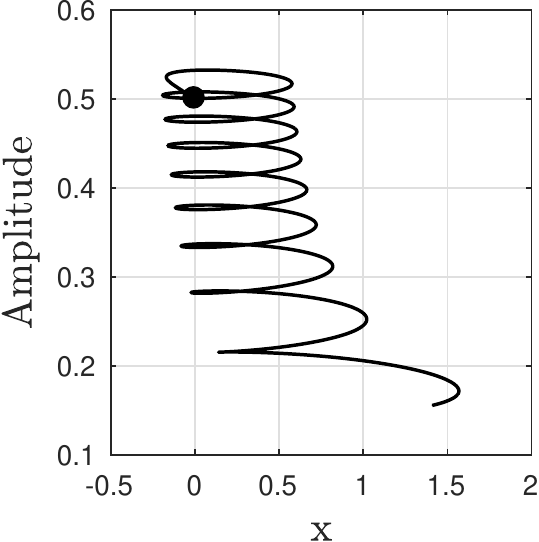}
         \includegraphics[scale =1]{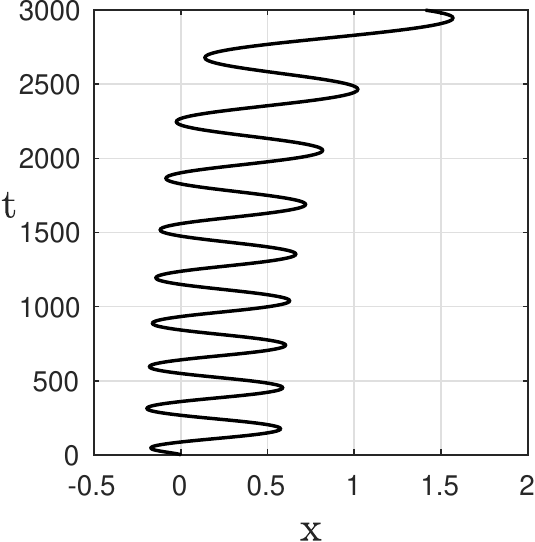}
	\caption{ Trapped soliton with a forcing with variable speed (\ref{lineardeviation}). Bottom (left): the amplitude of the trapped wave vs. its crests position. Bottom (right) the crest position of the soliton as a function of time. Parameters: $a=0.5$, $\Delta_0=a^{2}=0.25$, $g=-0.001$ and $b=1$.}
	\label{Fig4}
\end{figure}

\begin{figure}[h!]
	\centering	
	\includegraphics[scale =1]{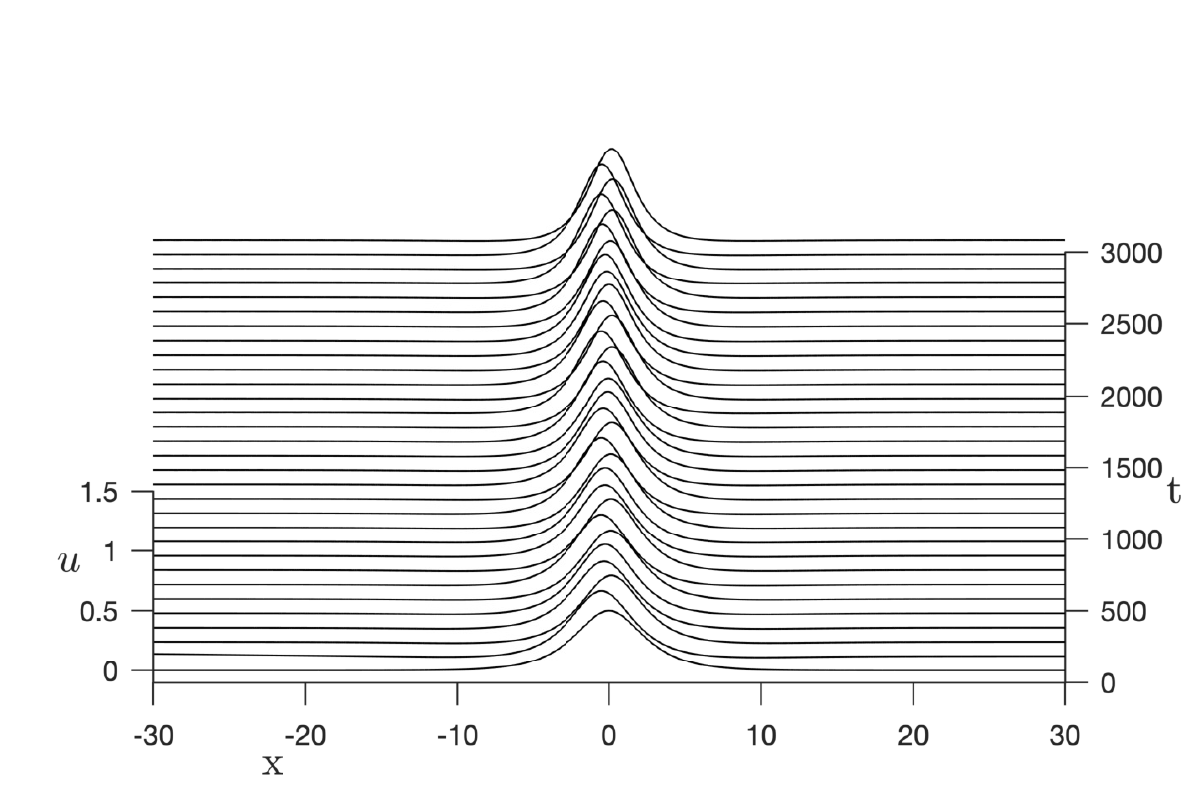}
 	\includegraphics[scale =1]{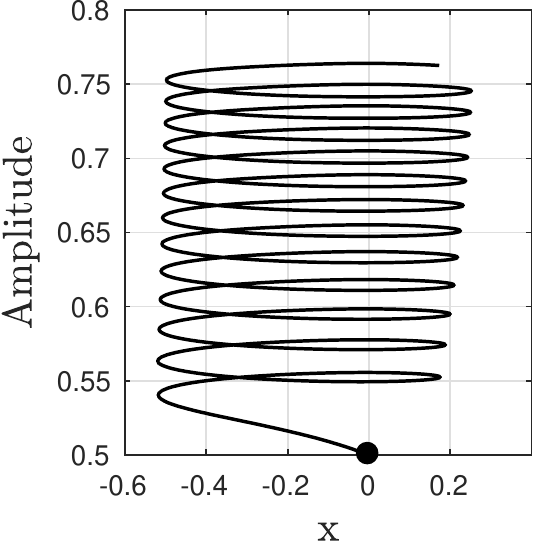}
         \includegraphics[scale =1]{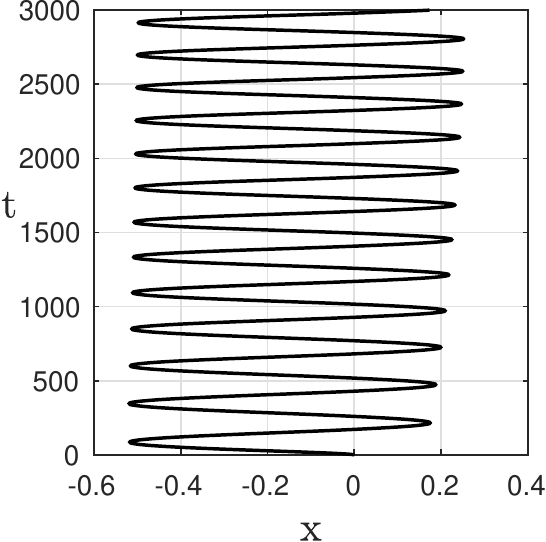}
	\caption{ Trapped soliton with a forcing with variable speed (\ref{lineardeviation}). Bottom (left): the amplitude of the trapped wave vs. its crests position. Bottom (right) the crest position of the soliton as a function of time. Parameters: $a=0.5$, $\Delta_0=a^{2}=0.25$ $g=0.001$ and $b=1$.}
	\label{Fig5}
\end{figure}

Results are in qualitative agreement with the asymptotic theory when the acceleration of the force is small.  More precisely, we have that for positive and small values of $g$ the amplitude of the soliton oscillates increasing with time, while for negative values of $g$ the soliton oscillates decreasing with time. This behaviour    resembles a spiral in the amplitude vs. crest position space. More details can be seen in Figure \ref{Fig4} and Figure \ref{Fig5}. As we can see in Figure \ref{Fig4} the soliton moves back and forth at the external forcing while its amplitude decreases over the time. Its amplitude gets close to zero and eventually the soliton structure vanishes. It is notable  that the soliton amplitude decays oscillating with time  without the presence of dissipative terms in the equation. A similar behaviour was reported in the recent work of Dinvay et al. \cite{Dinvay:2019} in the context of n moving loads on ice. The authors showed that an accelerated (or decelerated) load may lead to nonlinear superposition of trailing waves which may lead to ice break-up. Meanwhile, in Figure \ref{Fig5} we observe that the soliton is trapped at the external force, but its amplitude increases oscillation in time. Additionally, the oscillations in the space crest position vs. time presents an increasing in the oscillation, which resembles a forced harmonic oscillator. Although we cannot run simulations for all times because of computational limitations, this behaviour indicates that the soliton eventually escapes out of the external forcing.

\section{Conclusion}
In this paper, we have investigated asymptotically and numerically the interaction of solitons and a weak external field. We found that solitons can remain trapped at the external force for large times and regimes or  they move away from the external field without reversing its direction when the external field moves with constant speed.  In addition, when the speed of the external forcing varies linearly on time, we found that the amplitude of soliton can increase or decrease depending on wether the external forcing is accelerated of decelerated. These results agree qualitatively well  with the asymptotic theory for solitons of large amplitude.  The asymptotic theory predicts the polarity of the solitons may change during the interaction of the external force which does not occur in the fully numerical simulations.

\section{Acknowledgements}
Results described in Sect. 3a were obtained with support of RSF grant 22-17-00153.

	\section*{Declarations}
	
	\subsection*{Conflict of interest}
	The authors state that there is no conflict of interest. 
	\subsection*{Data availability}
	
	Data sharing is not applicable to this article as all parameters used in the numerical experiments are informed in this paper.

\end{document}